\documentclass[aps,prl,twocolumn,floatfix,superscriptaddress,showpacs]{revtex4}

\usepackage{graphicx}
\usepackage{bbm}
\usepackage{amsmath,amssymb}

\newcommand{\be}{\begin{equation}}
\newcommand{\beq}{\begin{equation}}
\newcommand{\ee}{\end{equation}}
\newcommand{\bea}{\begin{eqnarray}}
\newcommand{\eea}{\end{eqnarray}}
\newcommand{\ba}{\begin{array}}
\newcommand{\ea}{\end{array}}

\renewcommand{\vr} {{\bf r}}

\begin{document}

\title{Lower Bounds on the Exchange-Correlation Energy in Reduced Dimensions}
\author{E. R\"as\"anen}
\affiliation{Nanoscience Center, Department of Physics, University of
  Jyv\"askyl\"a, FI-40014 Jyv\"askyl\"a, Finland}
\affiliation{Institut f{\"u}r Theoretische Physik, Freie Universit{\"a}t Berlin,
Arnimallee 14, D-14195 Berlin, Germany}
\affiliation{European Theoretical Spectroscopy Facility (ETSF)}
\author{S. Pittalis}
\affiliation{Institut f{\"u}r Theoretische Physik, Freie Universit{\"a}t Berlin,
Arnimallee 14, D-14195 Berlin, Germany}
\affiliation{European Theoretical Spectroscopy Facility (ETSF)}
\author{K. Capelle}
\affiliation{Institut f{\"u}r Theoretische Physik, Freie Universit{\"a}t Berlin,
Arnimallee 14, D-14195 Berlin, Germany}
\affiliation{Departamento de F\'isica e Inform\'atica, Instituto de F\'isica de S\~ao
Carlos, Universidade de S\~ao Paulo, Caixa Postal 369, S\~ao Carlos, S\~ao
Paulo 13560-970, Brazil}
\author{C. R. Proetto}
\altaffiliation[Permanent address: ]{Centro At\'omico Bariloche and Instituto Balseiro,
8400 S. C. de Bariloche, R\'{i}o Negro, Argentina} 
\affiliation{Institut f{\"u}r Theoretische Physik, Freie Universit{\"a}t Berlin,
Arnimallee 14, D-14195 Berlin, Germany}
\affiliation{European Theoretical Spectroscopy Facility (ETSF)}

\date{\today}

\begin{abstract}
Bounds on the exchange-correlation energy of many-electron
systems are derived and tested. By using universal scaling properties of the 
electron-electron interaction, we obtain the exponent of the bounds in three, 
two, one, and quasi-one dimensions. From the properties of
the electron gas in the dilute regime, the tightest estimate to date is given 
for the numerical prefactor of the bound, which is crucial in practical 
applications. Numerical tests on various low-dimensional systems are in line 
with the bounds obtained, and give evidence of an interesting dimensional
crossover between two and one dimensions.
\end{abstract}



\maketitle

In 1979 Lieb \cite{lieb79} planted a landmark in quantum many-body physics by
proving the existence of a lower bound on the indirect part of the Coulomb 
interaction. The existence of such a bound is of immediate relevance to such 
fundamental questions as the stability of matter \cite{spruch91}.
For the purpose of quantitative calculations, on the other hand, existence
of a bound is not enough -- one would wish it to be as tight as possible.
A tighter version of Lieb's bound was later derived by
Lieb and Oxford \cite{lieboxford81}, and it is this tighter form,
known as Lieb-Oxford (LO) bound, which is used as key constraint in the 
construction of many modern density functionals \cite{pbe96,tpss04}, which
in turn are used in calculations of the electronic structure of atoms, 
molecules, nanoscale systems, and solids. 

In connection with recent advances in {\em low-dimensional} physics 
it is important to ask whether LO-like bounds exist and
can be formulated also in reduced dimensions, in particular since
the study of low-dimensional systems today forms a significant part of
condensed-matter and materials physics.

The LO bound~\cite{lieboxford81}, in its original form, applies to all 
{\em three-dimensional} (3D) nonrelativistic, Coulomb-interacting systems. 
The bound can be expressed in terms of the 
indirect part of the interaction energy 
\cite{lieb79,lieboxford81,levyperdew93},
\begin{equation}
W_{xc}[\Psi]\equiv \left<\Psi|{\hat V}_{ee}|\Psi\right>-U[n]\geq - \; C_3
\int d^3 r\,n^{4/3}(\vr),
\label{lo}
\end{equation}
where the electron-electron (e-e) interaction operator is Coulombic, i.e., 
${\hat V}_{ee}=\sum_{i>j}|\vr_i-\vr_j|^{-1}$. Its
expectation value is calculated over \textit{any}
normalized many-body wavefunction $\Psi (\mathbf{r}_{1},...,\mathbf{r}_{N})$.
$n(\mathbf{r}) $ is the corresponding density, and $U[n]$ is the classical Hartree energy.
For the prefactor $C_3$, where the subscript denotes the number of
dimensions $D$, Lieb originally found $C_3^{\rm L}=8.52$, which was subsequently refined by 
Lieb and Oxford to $C^{\rm LO}_3=1.68$, and later, numerically,
by Chan and Handy to $C^{\rm CH}_3=1.64$ ~\cite{chanhandy99}.
Recent numerical studies~\cite{odashimacapelle07,odashimacapelle08}, 
as well as modeling of the prefactor based on its known 
properties~\cite{odashimatrickeycapelle08}, have given evidence that 
the bound can be further tightened. 

In two-dimensions (2D), Lieb, Solojev and Yngvason~\cite{liebsolovejyngvason} 
(LSY) showed that
\be
W_{xc}[\Psi]\geq - \; C_{\rm 2} \int d^2 r\,n^{3/2}(\vr),
\label{2dbound}
\ee
where $C^{\rm LSY}_2=192\sqrt{2\pi}\approx 481 \gg C_3^{\text{LO}}$.
For a $D$-dimensional system, the bound 
may be written as
\begin{equation}
W_{xc}[\Psi]\geq - \; C_D \int d^D r\,n^\alpha(\vr),
\label{lod}
\end{equation}
but we note that the existence of a bound of this
form has been rigorously proven for only 3D and 2D, and that 
the tightest possible form ({\em i.e.}, the smallest possible value of
$C_D$) is unknown in all dimensions.

In this paper we (i) show that the exponents of $n$ in Eqs.~(\ref{lo}) and 
(\ref{2dbound}) are consequences of {\em universal} scaling properties of 
the e-e interaction; (ii) use this result to deduce the exponent 
$\alpha$ of a possible one-dimensional (1D) bound; (iii) provide an estimate of the prefactor $C_D$ 
that corresponds to a dramatic tightening of $C^{\rm LSY}_2$, smaller but
still significant tightening of $C^{\rm LO}_3$, and the first proposal for 
$C_1$; (iv) observe unexpected parameter independence and generality of 
the bound with respect to the model chosen for interactions in 1D; and (v) 
test the 1D and 2D bounds against analytical and near-exact numerical data 
for various low-dimensional systems.

The 1D case, in fact, is subtle because the Coulomb interaction is ill-defined.
Hence, we consider a contact interaction, 
${\hat V}_{ee}=\eta\sum_{i>j}\delta(x_i-x_j)$ with $\eta>0$. The discussion 
below on the 1D case refers to this type of interaction. However, we consider 
also a soft-Coulomb interaction, 
${\hat V}_{ee}=\sum_{i>j}\left[(x_i-x_j)^2+a^2\right]$,
which corresponds to a quasi-1D (q1D) situation. 

Under homogeneous scaling of the coordinates, $\vr\rightarrow \gamma\vr$ 
($0<\gamma<\infty$)~\cite{levyperdew93}, the $(DN)$-dimensional
many-body wavefunction scales as $\Psi(\vr_1\ldots \vr_N)\rightarrow 
\Psi_\gamma(\vr_1\ldots \vr_N)=\gamma^{DN/2}
\Psi(\gamma\vr_1\ldots \gamma\vr_N)$,
preserving normalization. This yields the number-conserving scaled density
$n(\vr)\rightarrow n_\gamma(\vr)=\gamma^D n(\gamma\vr)$.
On the other hand, $W_{xc}[\Psi] \rightarrow W_{xc}[\Psi_{\gamma}]
= \gamma W_{xc}[\Psi]$,
since both the Coulomb ($D=2, 3$) and contact ($D=1$) interaction, 
and their Hartree approximations scale linearly. 
Thus, Eq.~(\ref{lod}) becomes
\begin{equation}
\gamma W_{xc}[\Psi]\geq - \; C_D \; \gamma^{D(\alpha-1)} \int d^D r\,n^\alpha(\vr),
\label{lodg}
\end{equation}
and consistency between Eqs.~(\ref{lod}) and (\ref{lodg}) immediately
yields $\gamma=\gamma^{D(\alpha-1)}$, giving $\alpha=1/D+1$. For $D=3$ and 
$D=2$ this yields $\alpha=4/3$ and $\alpha=3/2$, respectively, in 
agreement with the LO and LSY bounds. We thus find that if a bound of this form
exists, its exponent is, in all dimensions, uniquely determined by 
coordinate scaling, without requiring the complicated ana\-ly\-sis performed in 
Refs.~\cite{lieb79}, \cite{lieboxford81}, and \cite{liebsolovejyngvason}.
For a LO-like bound in 1D, the same scaling argument suggests the form
$W_{xc}[\Psi]\geq - \; C_1 \int d^1 r\,n^{2}(\vr)$,
although the existence of such a bound in 1D is at present only a conjecture.

The exponent $\alpha=1/D+1$ in Eq.~(\ref{lodg}) is the same as in the 
expression of the exchange energy, $E_x[n]$, of the homogeneous $D$-dimensional
electron gas, which is applied in the local-density approximation 
(LDA)~\cite{GV} to the inhomogeneous case. Thus, we can express the right-hand 
side of all LO-like bounds in terms of
\be
E_x^{\rm LDA}[n] = - \; A_D \int d^D r\,n^\alpha(\vr),
\label{lda}
\ee
where $A_3=3^{4/3}\pi^{-1/3}/4$, $A_2=2^{5/2}\pi^{-1/2}/3$, and $A_1=\eta/4$ 
~\cite {GV}. For ground-state densities, the left-hand-side of Eq.~(\ref{lod}) 
can be written in terms 
of the full exchange-correlation energy, $E_{xc}[n]\equiv W_{xc}[n]+T_c[n]\geq W_{xc}[n]$,
where the inequality follows from the positiveness of the kinetic-energy
part $T_c$ of the correlation energy, and the density functional
$W_{xc}[n]$ is obtained by evaluating $W_{xc}[\Psi]$ with the wavefunction 
minimizing
$\left\langle \Psi \left\vert \widehat{T}+ \widehat{V}_{ee}\right\vert \Psi \right\rangle $ 
under the constraint of reproducing the ground-state density $n(\mathbf{r})$.
We can now cast the bound in the form 
\be
E_{xc}[n]\geq \lambda_D \; E_x^{\rm LDA}[n],
\label{const_lambda}
\ee
where $\lambda_D=C_D/A_D$. Thus, the tightest possible bound
can be obtained by
looking for the {\em maximum} value of the density functional
\be
\lambda_D[n] = \frac{E_{xc}[n]}{E_x^{\rm LDA}[n]}
=\frac{E_{x}[n]}{E_x^{\rm LDA}[n]} + \frac{E_{c}[n]}{E_x^{\rm LDA}[n]}
\label{LObound}
\ee
over all possible $D$-dimensional many-body systems.
Clearly, this maximization cannot be performed in practice. However, we can 
make an educated guess as to what the resulting maximum will be. 

First, we
note that for 3D, it was shown rigorously that the constant $\lambda_3$
in Eq.~(\ref{const_lambda}) 
can be replaced by a monotonic function depending on particle number, 
$\lambda_3(N)$, which assigns to all systems with particle number $N$ a 
common value $\lambda_3(N)\leq \lambda_3(N\to\infty)\equiv \lambda_3$, 
such that a LO-like bound with $\lambda(N)$ in place of $\lambda$ holds 
for all systems with this $N$ \cite{lieboxford81,odashimatrickeycapelle08}. 
The values commonly quoted for $\lambda_3^{\rm L}$, $\lambda_3^{\rm LO}$, 
and $\lambda_3^{\rm CH}$ are actually estimates of $\lambda_3(N\to\infty)$.
To obtain the tightest possible universal ($N$-independent) bound we thus 
take $N\to \infty$. 

Second, we may expect that $E_{x}$ is always relatively close to 
$E_x^{\rm LDA}$, as the particle density $n$ is varied over different physical 
systems. Hence, the functional $\lambda_D[n]$ is expected to be largest for 
systems where the rightmost ratio in Eq.~(\ref{LObound}) is largest, 
{\em i.e.}, where correlation is largest relative to exchange. This situation 
is typical of the extreme low-density limit $n\to 0$. 

Taken together, the rigorous property of a maximum at $N\to\infty$, and
the nonrigorous but reasonable requirement that $n\to 0$,
suggest that the largest possible value of $\lambda_3[n]$ is obtained
for the $r^{\rm 3D}_s\rightarrow\infty$ limit of the 3D 
electron gas (3DEG) with the density parameter 
$r^{\rm 3D}_s=3^{1/3}(4\pi n)^{-1/3}$~\cite{perdewwang}. This expectation leads to
$\lambda_3 \equiv \lambda_{\rm 3DEG}[r^{\rm 3D}_s\rightarrow\infty]=
1+\epsilon_c(r^{\rm 3D}_s\rightarrow\infty)/\epsilon_x(r^{\rm 3D}_s\rightarrow\infty)
=1.9555$~\cite{odashimacapelle07}, where $\epsilon_x$ ($\epsilon_c$) is 
the exchange (correlation) energy per electron.
We note that the original LO bound with $C_3^{\rm LO}=1.68$
corresponds to $\lambda_3^{\rm LO}=2.27$, while the present
estimate $\lambda_3=1.9555$ is tighter, and consistent with the 
empirical prefactor obtained by evaluating 
$\lambda_3[n]$ for real systems ~\cite{odashimacapelle07}.

We now assume that the above argument about the maximum of $\lambda_3$ 
carries over to reduced dimensions. For the 2D electron gas (2DEG)
with $r_s^{\rm 2D}=1/\sqrt{\pi n}$ we have 
$\epsilon_x(r_s^{\rm 2D})=-4\sqrt{2}/(3\pi r^{\rm 2D}_s)$.
From the Madelung energy of the Wigner crystal~\cite{bonsall}
we extract that the 
leading contribution to the correlation in the 
low-density limit is
$\epsilon_c(r_s^{\rm 2D}\rightarrow\infty)=-0.509/r_s^{\rm 2D}
+0.815/\left( r_{s}^{\rm 2D}\right) ^{3/2}$. 
Thus we find 
$\lambda_2 \equiv \lambda_{\rm 2DEG}[r^{\rm 2D}_s\rightarrow\infty]=1.84$,
which is a dramatic improvement on the rigorous mathematical result
of Ref.~\cite{liebsolovejyngvason} that 
$\lambda_2\leq \lambda_2^{\rm LSY}\approx 452$.

Analogously, for the 1D electron gas (1DEG) with a contact interaction
and $r_s^{\rm 1D}=1/2 n$ we find 
$\epsilon_x(r_s^{\rm 1D})=-\eta/(8 r^{\rm 1D}_s)$.
The leading contribution to the correlation energy in the low-density limit is 
$\epsilon_c(r_s^{\rm 1D}\rightarrow\infty)=
-\eta/(8 r_s^{\rm 1D})+\pi^2/32(r_{s}^{\rm 1D})^{2}$~\cite{magyarburke04}, 
yielding $\lambda_1 \equiv \lambda_{\rm 1DEG}[r^{\rm 1D}_s\rightarrow\infty]=2$.
Note that this is a {\em general} result, in the sense that it does not depend
on the parameter $\eta$ of the contact interaction.

Finally, in the q1D system with a soft-Coulomb interaction
we first note that the scaling of $W_{xc}$
is nontrivial due additional scaling of the parameter $a$: 
$W_{xc}^a[\Psi] \rightarrow W_{xc}^a[\Psi_{\gamma}] = 
\gamma W_{xc}^{\gamma a}[\Psi]$.
Second, for the LDA in Eq.~(\ref{lda}) we have now $A_{\text{q1D}}=1/2$,
$\alpha=2$, and in the limit $a\rightarrow 0$ the
integrand is multiplied by a function 
$f^a[n(x)]=\ln[2/\pi a n(x)]+3/2-\mu$, where $\mu\simeq 0.577$ is Euler's
constant~\cite{fogler}. Note the scaling
property $f^a[n(x)] \rightarrow f^a[n_{\gamma}(x)]= \gamma f^{\gamma a}[n(x)]$.
Now, we assume that the LO bound in q1D also has this
form, {\em i.e.}, the integrand of the q1D expression for
$W_{xc}$ [Eq.~(\ref{lod})] is multiplied by the
same factor $f^a[n(x)]$. Under this assumption, we may
search for the maximum values for $\lambda_{\rm q1D}[n]$ in 
Eq.~(\ref{LObound}) in a similar fashion as in 3D, 2D, and 1D. 

In the low-density limit of the q1D electron gas (q1DEG) we have
$\epsilon_x(r_s^{\rm q1D}\rightarrow\infty)=n[\ln(a\pi n/2)-3/2+\mu]/2$
and $\epsilon_c(r_s^{\rm q1D}\rightarrow\infty)=n[\ln(a n/2\pi)+3/2+\mu]/2$
\cite{fogler}. Hence, we find 
$\lambda_{\rm q1D} \equiv \lambda_{\rm q1DEG}[r^{\rm
  q1D}_s\rightarrow\infty]=2$.
We note, again, that the leading contribution to $\lambda_{\rm q1DEG}$ is 
{\em independent} of the softening parameter $a$ of the q1D model.

Moreover, we note the highly nontrivial fact that the leading contribution to
$\lambda_{\rm 1DEG}$ and $\lambda_{\rm q1DEG}$ is the {\em same}. 
This encourages us to propose $\lambda_D=2$ as the tightest general
bound for both 1D and q1D. It is also important to note that in 
2D, 1D, and q1D, the correction to the leading term is negative,
decreasing the value of the corresponding $\lambda_D$ for 
finite (non-vanishing) 
densities, in line with our proposal to extract the maximum of $\lambda_D[n]$ 
from the low-density limit of the electron gas.
The results for the tightest bounds in different dimensions 
are summarized in Table~\ref{table}.
\begin{table}
  \caption{\label{table} 
Estimated prefactors $\lambda_D = C_D/A_D$ for bounds on $E_{xc}$ and
$W_{xc}$, compared to literature values, in different dimensions. 
The quasi-one-dimensional (q1) case involves special conditions (see text).
}
  \begin{tabular}{c | c c c c}
  \hline
  \hline
  $D$    & 3 & 2 & 1 & q1 \\
$\lambda_D^{\rm here}$ & \;\; 1.96 \;\; & \;\; 1.84 \;\; & \;\; 2.0 \;\; & \;\; 2.0 \;\;\\ 
$\lambda_D^{\rm lit}$ & 2.27~\cite{lieboxford81} & 452~\cite{liebsolovejyngvason} & - & -\\ 
  \hline
  \hline
  \end{tabular}
\end{table}

Next, we test our bounds against analytical and near-exact numerical 
data obtained independently for low-dimensional systems, 
in a similar spirit as was done for 3D systems in 
Ref.~\cite{odashimacapelle07}.
In particular, we consider 2D parabolic (harmonically confined) and 
hard-wall square quantum dots~\cite{qd-reviews} 
(QDs), where the density parameter can be estimated
as $r_s=N^{-1/6}\omega^{-2/3}$ (Ref.~\cite{koskinen}) and $r_s=(\pi
N)^{-1/2}L$ (Ref.~\cite{polygonal}), respectively (2D indication
omitted for clarity). Here $\omega$ is the
harmonic confinement strength, $L$ is the side length of the square
QD, and $N$ is the number of electrons.

\begin{figure}
\includegraphics[width=0.99\columnwidth]{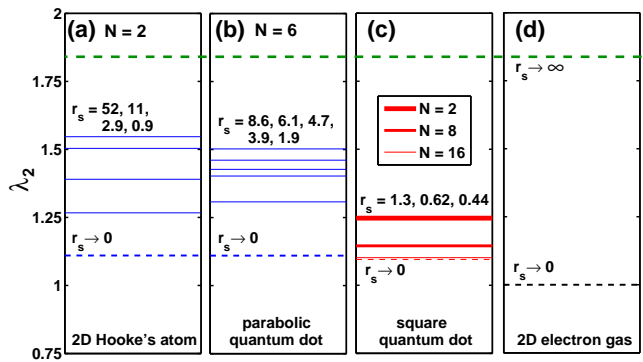}
\caption{(Color online) Values for $\lambda_2[n]$ in different 
finite and infinite 2D systems compared to our limit $\lambda_2=1.84$
(thick dashed line).}
\label{panels}
\end{figure}
Figure~\ref{panels}(a)
shows $\lambda_2[n]$ for a 2D Hooke's atom, which is equivalent to a
parabolic QD with $N=2$. Here we focus on some of the analytical
two-electron solutions in the range $r_s=0.9\ldots 52$ 
derived by Taut~\cite{taut}. The maximum value 
$\max(\lambda_{2,{\rm Hooke}})\approx 1.55$ is relatively close 
to the corresponding 3D result $\max(\lambda_{3,{\rm Hooke}})\approx 1.49$ 
(Ref.~\cite{odashimacapelle07}).
Detailed analysis of the low-density behavior of $\lambda_2$ 
is given below. In the noninteracting (high-density) limit,
where the correlation energy is zero, we find numerically 
$\lambda_{2,{\rm Hooke}}^{\omega\rightarrow\infty}=E_x/E_x^{\mathrm{LDA}}\approx 1.10$
which is slightly below the corresponding 3D
value~\cite{odashimacapelle07} of 1.17.

In larger QD systems ($N>2$), most reference data are given only in terms of
ground-state total energies $E_{\mathrm{tot}}$, whereas the calculation of 
$\lambda_D[n]$ requires knowledge of the exact exchange-correlation 
energy and the electron
density $n$. The exact DFT correlation energy can be computed as 
$E_c[n^{\mathrm{exact}}] = E_{\mathrm{tot}}[n^{\mathrm{exact}}] - 
E_{\mathrm{tot}}^{\mathrm{EXX}}[n^{\mathrm{exact}}]$ where 
$E_{\mathrm{tot}}$ is
the exact total energy and EXX refers to exact exchange. To
estimate $\lambda_D[n^{\mathrm{exact}}]$, we may then perform a
self-consistent EXX calculation and calculate 
\begin{equation}
\lambda_D \approx \frac{E_x^{\mathrm{EXX}}[n^{\mathrm{EXX}}]+E_{\mathrm{tot}
}[n^{\mathrm{exact}}]-E_{\mathrm{tot}}^{\mathrm{EXX}}[n^{\mathrm{EXX}}]}
{E_x^{{\mathrm{LDA}}}[n^{\mathrm{EXX}}]}.  \label{lambda_app}
\end{equation}
In this work we have performed the EXX
calculations in the Krieger-Li-Iafrate (KLI) approximation~\cite{KLI}
within the {\tt octopus} real-space density-functional code~\cite{octopus}. 
We note
that according to our numerical test for the 2D Hooke's atom, the estimate
in Eq.~(\ref{lambda_app}) yields generally larger values for $\lambda_D[n]$ 
than the definition in Eq.~(\ref{LObound}).

Figure~\ref{panels}(b) shows results for a parabolic QD with $N=6$. Here
we use as the reference data the variational quantum Monte Carlo (QMC) 
total energies in the weak-confinement regime~\cite{ari_wigner}. 
In Fig.~\ref{panels}(c) we present results for square-well QDs with $L=\pi$
and varying $N$. Again, we use variational QMC
data for the total energies~\cite{recta}. Comparison with fixed $r_{s}$ to
parabolic -- and also to circular-well QDs (not shown) -- reveals that 
deformation from the circular geometry decreases $\lambda_2[n]$. 
Similar decrease in $\lambda_2[n]$ is found if the circular confinement
is made elliptic in the $r_{s}=0$ limit (not shown).

The $r_s=0$ limit allows testing also \emph{within} circular
confinement by varying the curvature, {\em i.e.}, the exponent in 
$V_{\mathrm{circular}}(r)=|r|^\alpha$. Interestingly, the largest value for 
$\lambda_2$ is obtained at the smallest $\alpha$ we can numerically consider, 
{\em i.e.}, at $\alpha=0.5$, which gives $\lambda_2[n]=1.110$. 
Overall, the numerical results summarized in Fig.~\ref{panels} show that
in both finite and infinite 2D systems, values obtained for $\lambda_2[n]$ 
are consistently below our limit $\lambda_2=1.84$ (thick dashed line).

\begin{figure}
\includegraphics[width=0.7\columnwidth]{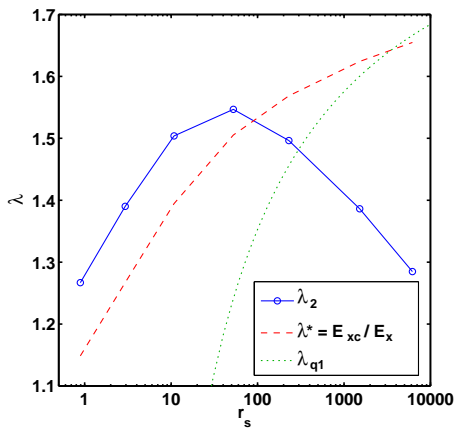}
\caption{(Color online) Values for $\lambda_2[n]$ in a two-dimensional
Hooke's atom as a function of $r_s=2^{-1/6}\omega^{-2/3}$. 
The circles connected by solid lines
correspond to $\lambda_2[n]$ in the definition of Eq.~(\ref{LObound}). 
The dashed line shows the auxiliary quantity $\lambda^*[n]$ (see text).
The dotted line corresponds to the quasi-one-dimensional 
result.}
\label{hookium}
\end{figure}

Finally we consider the low-density limit of the 2D Hooke's atom
in detail. In Fig.~\ref{hookium}
we show $\lambda_2[n]$ (solid line) up to the extreme low-density regime. 
As expected, $\lambda_2[n]$ first increases as a function of $r_s$.
However, at $r_s\sim 50$,
we find an local maximum of $\lambda_2[n]\approx 1.55$ 
[see also Fig.~\ref{panels}(a)] followed by a decrease
at higher $r_s$. By contrast, if the (2D) LDA exchange in
Eq.~(\ref{LObound}) is replaced by exact exchange,
$\lambda_2[n]\rightarrow\lambda^*[n]:=E_{\mathrm{xc}}/E_x=1+E_c/E_x$, 
the behavior is monotonic (dashed line) as expected. Note, however, that 
$\lambda^*[n]$ is not the quantity used in the LO bound, but is
used here as an auxiliary quantity.
Examination of the total electron density in the low-density regime suggests
that the unexpected behavior of $\lambda_2[n]$ in Fig.~\ref{hookium} 
is due to the breakdown of
the 2D-LDA. Namely, as the confinement is made
weaker, the electrons are pushed further apart from each other leading to a
q1D ring-shaped total density. In fact, 
the low-density regime in Fig.~\ref{hookium} shows 
reasonable agreement between $\lambda^*$ and the q1D result (dotted
line) deduced from Fogler~\cite{fogler} with the parameter
$a$ estimated from the low-density ring-like model by Taut~\cite{taut}.
Hence, it is evident that decreasing the density in a 2D Hooke's atom
leads to a dimensional crossover. 

To summarize, we have shown that the exponents in Eqs.~(\ref{lo}) 
and (\ref{2dbound}) are consequences of {\em universal} scaling properties of
the electron-electron interaction. We have thus been able to deduce the 
exponent $\alpha$ of a one-dimensional bound. Furthermore, we have 
provided a tightening of the prefactor of the 
three-dimensional bound,
a dramatic tightening of the prefactor in two-dimensions, and the
first proposal for the prefactor in one dimension. 
Unexpected generality of the bound with respect to the type of
interactions in one- and quasi-one-dimensional systems was observed. 
Our numerical tests for
low-dimensional model systems are consistent with the derivations, 
all showing $\lambda_2[n]<\lambda_2$, and
display an interesting dimensional crossover in the low-density limit.
Besides their general relevance in quantum many-body physics, 
these results provide constraints for accurate approximations of  
the exchange-correlation functionals in any dimension.

\begin{acknowledgments}
We thank M. Taut for providing us with numerical data of 
the two-dimensional Hooke's atom, and A. Harju for the
QMC data. This work was supported by the
Deutsche Forschungsgemeinschaft and the EC's 6th FP through
the Nanoquanta NoE (NMP4-CT-2004-500198). In addition, E. R. 
was supported by the Academy of Finland, C. R. P. by EC's Marie
Curie IIF (MIF1-CT-2006-040222), and K. C. by FAPESP and CNPq.
\end{acknowledgments}

\end{document}